# Possible Origin of the Anomalous Far-infrared Absorption by Small Metal Particles

Y. H. Kim[*]

Department of Physics, University of Cincinnati, Cincinnati, Ohio 45221, U. S. A.

and

Günter Schmid

Universität Duisburg-Essen, Institut für Anorganische Chemie, Essen 45117, Germany

We have carried out the far-infrared transmission measurements on the ligand stabilized $Au_{55}$ particles dispersed in CsI. We have found that there exists the crossover energy $E_c \sim 160$ cm$^{-1}$ (~ 19 meV) beyond which the absorption coefficient recovers the classical electric dipole-like absorption and, hence, the absorption for frequencies higher than $E_c$ is insensitive to the particle size. We propose that the physical origin of the enigmatic anomalous far-IR absorption by small metal particles lies in the absorption arising from the transitions the electrons between the discrete energy levels below $E_c$ on which the particle size independent, classical electric dipole absorption develops for frequencies higher than $E_c$.

[*]email address: kimy@ucmail.uc.edu

History of small metal particles goes back to the time of the discovery of the fascinating colors displayed in glasses that contain colloidal particles of gold or/and silver [1]. This optical phenomenon, which is known as the Mie resonance [2], has its origin in the oscillating electric dipole moment of small metal particles strongly coupled to electromagnetic fields. What is more fascinating is that the magnitude of the measured far-infrared (far-IR) absorption coefficient ($\alpha$) is orders of magnitude larger than the corresponding far-IR absorption tail of the Mie resonance which is given by $\alpha \approx 9 f \varepsilon_m^{3/2} \omega^2 / 4\pi c \sigma_1$ for small volume fraction $f$ of small metal particles of the real part of conductivity $\sigma_1$ dispersed in an insulating medium of dielectric constant $\varepsilon_m$ [3]. Even though it was realized that the additional far-IR absorption due to the induced magnetic dipole, which arises from the eddy current, given by $\alpha_M = 2\pi f \varepsilon_m^{1/2} a^2 \sigma_1 \omega^2 / 5 c^3$ becomes significant for larger size particles of radius $a$ [3], $\alpha_M$ alone cannot account for the discrepancy between the experimental $\alpha$ and that of the Mie resonance tail. In particular, the far-IR absorption by small metal particles whose radius is smaller than 50 Å where $\alpha_M$ becomes negligibly small, the measured far-IR absorption was still orders of magnitude larger than $\alpha$ of the electric dipole tail although the frequency dependence was the same.

The first far-infrared (far-IR) measurements of small Cu, Al, Sn, and Pb particles of diameter ranging from 65 Å to 350 Å by Tanner *et al*. [4] found that $\alpha$ was proportional to $\omega^2$ and anomalously larger than the calculated $\alpha$ by several orders of magnitude. This tendency has been consistently demonstrated to be true although the role of clustering of the metal particles was attributed to the origin of the enhancement of the far-IR absorption by several authors [5 - 7]. However, the latest study on the clustering effect and particle size dependence on the far-IR absorption by small aluminum particles [8] found that while clustering did enhance the absorption by a factor of 10, the magnitude of the absorption by the well-isolated aluminum particles was still larger than the classical electric dipole absorption by two orders of magnitude. Moreover, the measured absorption coefficient showed little particle size dependence [8].

When the metal particle size is smaller than the mean free path of the electrons in the bulk, quantum mechanical treatment of the far-IR absorption becomes necessary because of the discrete energy levels with the average energy gap $\overline{\Delta} = E_F/N$ if all the degeneracy is lifted [9]. Here $E_F$ is the Fermi energy the electrons and $N$ is the number of free electrons in the particle). In quantum regime the far-IR absorption comes from the transitions of the electrons between the energy levels, whose strengths are proportional to the dipole matrix element squared. Then the classical electric dipole absorption behavior would be recovered for frequencies higher than $\overline{\Delta}$ which sets the energy scale for crossover from a molecule to a solid. It was found that quantum mechanical calculation of the far-IR absorption agrees with classical theory when $\omega >> \overline{\Delta}$ as expected but a reduced absorption when $\omega \approx \overline{\Delta}$ [10, 11]. Thus it is concluded that the anomalous far-IR absorption cannot be explained as due to quantum size effect. Therefore the origin of the anomalous far-IR absorption still remains controversial, waiting for a definite answer.

The small metal particles used for the far-IR absorption studies were typically prepared by the gas-evaporation method that produced small metal particles of radius typically in the range between 20 Å and 400 Å in the form of smoke [12]. One drawback of this technique was the difficulty in the control of the particle size and of the particle size distribution. This particle size distribution has been attributed to the lack of the experimental observation of the size quantized energy gap. For the far-IR absorption measurements the small metal particle smoke was dispersed in a far-IR transparent insulating medium such as alkali halides, polyethylene or Teflon at a desired volume fraction. The smoke and dielectric medium mixture was mixed in a freezer mill operating at 77 K and then pressed into a pellet. Regrinding and compressing cycle of the pellet for several times was necessary to ensure the uniform distribution of isolated particles. However, there was no sure way of controlling the clustering of the small metal particles through the grinding/compressing technique.

The breakthrough occurred when the ligand stabilized $Au_{55}$ particles ($Au_{55}(PPh_3)_{12}Cl_6$) was chemically synthesized [13]. $Au_{55}(PPh_3)_{12}Cl_6$ nanoparticle (Au55 hereafter) has a diameter 14 Å surrounded by $PPh_3$ molecules and Cl atoms, making the size of Au55 be 21 Å in diameter. Since the Au55 nanoparticles are chemically synthesized just as any other molecules, their sizes are identical. Furthermore the presence of the ligand ($PPh_3$ = Triphenylphosphine) molecules on each Au55 particle guarantees electrical isolation between the particles. Therefore, we have a perfect window of opportunity to look into the outstanding critical issues of the quantum size effect and the anomalous far-IR electric dipole absorption.

In this work, we have carried out the far-IR absorption measurements on Au55 particles dispersed in CsI as an insulating medium which is known to be transparent for frequencies above 100 cm$^{-1}$ in order to find the spectral information that was not accessible with Teflon. The Au55 particle/CsI mixture (Au55/CsI) at $f = 0.002$ was ground at 77 K in order to ensure uniform distribution of Au55 particles in the CsI matrix. Then the Au55/CsI mixture was compressed into a 1 cm diameter pellet until CsI flows. The pressing/regrinding was repeated 3 cycles. The pellet thickness was maintained at around ~ 3 mm. The far-IR transmission measurements were done using a Bruker 113v spectrometer in conjunction with a Si-composite bolometer detector normally operating at 4.2 K. In order to cover the frequency between 100 cm$^{-1}$ and 500 cm$^{-1}$, 6 µm thick Mylar beam splitter was used. For mid-IR range, Ge/KBr beam splitter was used. Far-IR absorption coefficient was calculated from the measured transmission ($T$) via $\alpha(\nu) \approx -\ln(T)/d + 2\ln(1-R)/d$ for small $R$ where $R$ is the far-IR reflectivity and $d$ is the thickness of the sample. Here $\nu$ is the frequency measured in cm$^{-1}$. The absorption coefficient of Au55 particles, $\alpha_{Au55}$ was then calculated by subtracting the absorption coefficient of CsI, $\alpha_{CsI}$ from $\alpha_{Au55/CsI}$ by assuming that $R_{Au55/CsI} \approx R_{CsI}$ for small $f$. $\alpha_{Au55}$ for frequencies between 100 cm$^{-1}$ and 500 cm$^{-1}$ is displayed in Figure 1 in black along with the far-IR absorption coefficient of Au55 in Teflon for $\nu < 160$ cm$^{-1}$ in gray. In order for direct comparison with the far-IR absorption of the Au55 particles dispersed in Teflon [14], the far-IR absorption coefficient was rescaled to match the $f = 0.01$ Au55/Teflon absorption coefficient since the far-IR absorption is linear in $f$ for a dilute small metal particle system [3, 14].

The classical electric dipole absorption was calculated by modeling the Au55 particle as a sphere that contains 37 electrons (55 valence electrons less 18 electrons that are spent to form chemical bonding with the ligand molecules and Cl atoms) at the density of bulk Au at $n_{Au} = 5.90 \times 10^{22}$ cm$^{-3}$, which corresponds to the particle radius $a \sim 5.3$ Å. Then $\sigma_1$ of Au55 may be written as $\sigma_1 = 11.89 a$ $\Omega^{-1}$cm$^{-1}$ with $a$ in Å using the Fermi velocity for Au $v_F = 1.4 \times 10^{10}$ cm/s for the scattering rate $\Gamma = a/v_F$ ($a$ is smaller than the electron mean-free-path in bulk Au) and, therefore, $\alpha_{Au} = 0.08 f \nu^2 / a$. It turned out that CsI had weak absorption features at $\sim 220$ cm$^{-1}$ and $\sim 420$ cm$^{-1}$ as also shown in Figure 1 in light gray. Note that while the absorption feature of CsI at $\sim 220$ cm$^{-1}$ $\alpha_{Au55/CsI}$ is almost subtracted out, the CsI mode at $\sim 420$ cm$^{-1}$ in $\alpha_{Au55/CsI}$ appears overcompensated during the subtraction of $\alpha_{CsI}$ from $\alpha_{Au55/CsI}$ because the absorption due to 420 cm$^{-1}$ mode in $\alpha_{Au55/CsI}$ is less than that of $\alpha_{CsI}$.

The measured far-IR absorption behavior does not resemble that of the classical electric dipole absorption of $a = 5.3$ Å Au55 particles (dotted line in Figure 1) at all. Rather it follows $\alpha \sim 0.11(\nu - 10)^{0.8}$ (dashed line) and flattens out for $\nu \geq \sim 160$ cm$^{-1}$. The far-IR absorption by Au55 particles is several times larger than that of the electric dipole absorption and its frequency dependence $\sim (\nu - \Delta)^{0.8}$ is closer to that of a three-dimensional (3D) semiconductor of an energy gap $\Delta \sim 10$ cm$^{-1}$ with an exponent 0.8 instead of 0.5. The magnitude of the measured absorption of Au55 particles becomes comparable to that of the classical electric dipole absorption at $\nu \sim 160$ cm$^{-1}$ indicated as $E_c$ beyond which the absorption stays flat (see Figure 1).

This illustrates that the electrons in the small metal particle no longer "see" the surface boundary for $\nu > E_c$, and enter the classical regime where the electric dipole absorption sets in, making the discreteness of the electron energy level become irrelevant and the electric dipole absorption be independent of the size of the particle. This $E_c$ is physically equivalent to the ionization energy in the Bohr model for hydrogen atom beyond which electrons enter into the continuum in energy state. Therefore, by taking the effective Bohr radius ($a_B$) of the electrons in a Au55 particle as the particle size of 14 Å, one finds $E_c = m_e e^4 / 2\varepsilon_s^2 \hbar^2 = \hbar^2 / 2 m_e a_B^2 \sim 0.019$ eV ($\sim 160$ cm$^{-1}$). Here $\varepsilon_s$ is the dielectric constant responsible for the screening of the electrons in Au55 which can also be estimated as $\varepsilon_s = a_B m_e e^2 / \hbar^2 \sim 27$. This value is a typical value that comes from the contribution of atomic and ion remainder shells in solids. Thus poorer screening of the electrons in Au55 is expected in comparison with its bulk state.

When the IR frequency is increased further into mid-IR as high as $\sim 0.5$ V, it is evident that $\alpha_{Au55}$ recovers the classical behavior marked by $\nu^2$–dependence in the absorption coefficient curve as shown in Figure 2. In addition there are a number of PPh3 related IR modes for 500 cm$^{-1}$ < $\nu$ < 2000 cm$^{-1}$ and C-H stretch mode at $\sim 3000$ cm$^{-1}$, and O-H stretch modes near 3500 cm$^{-1}$ which is probably due to the hygroscopic nature of CsI. The rest of the modes except for the O-H stretch modes belong to the PPh$_3$ ligand molecules [15]. The derivative-like shape of the modes that appear in $\alpha_{Au55}$ even after subtracting the pure ligand absorption $\alpha_{Ligand/CsI}$ are due to the different bonding states between the free PPh$_3$ and the covalently bonded PPh$_3$ to Au$_{55}$ particle. However, the $\nu^2$-dependent background of $\alpha_{Au55}$ is readily seen.

As displayed in Figure 3, the classical electric dipole absorption for $a = 5.3$ Å (dotted curve) shows entirely different frequency dependence and its magnitude is much less that the observed absorption coefficient for $\nu < E_c$. Interestingly a reasonable fit to the experimental data of Au55 particle was obtained with an unphysical $a = 400$ Å chosen to match the overall magnitude of the absorption with a constant 4 cm$^{-1}$ offset. Since the electrons in small metal particle do not suffer the scattering from the surface anymore for $\nu > E_c$, the corresponding absorption should be particle size independent. In other words, the surface boundary condition does not dictate the dynamics of the electrons for frequencies above $E_c$. This bears a significant physical meaning that might hold the key to the answer to the longstanding puzzle, namely the anomalous far-IR absorption by small metal particles. This observation implies that the magnitude of the far-IR absorption is offset by the absorption in the quantum regime set by the energy scale $E_c$ not $\Delta$. Then the far-IR absorption strength of the transitions between the discrete energy levels must be drawn from the strength of the Mie resonance as the sum rule argument suggests. In fact the studies of the surface plasmon absorption cross-section of Na clusters, it was found that there was missing oscillator strength up to 25% [16]. It was conjectured that at least 50% of the missing strength should go to the far-IR [17].

Since the $E_c$ of all the small metal particles (20 Å $< a <$ 400 Å) studied in the past falls into the millimeter to microwave range, all the observed far-IR absorption in the 10 cm$^{-1}$ – 100 cm$^{-1}$ window would appear anomalously large compared with the classical electric dipole absorption because of its tie to $E_c$. In other words, the larger the particle size is, the smaller $E_c$ is, and the larger the measured far-IR absorption coefficient becomes. This observation also explains the weak particle size dependence of the measured far-IR absorption coefficient compared with that of the classical electric dipole absorption [8]. Additional magnetic dipole contribution to absorption could also give the particle size dependence ($\propto a^2$) for larger size particles.

In summary, we have found that there is a crossover energy above which the absorption coefficient of Au55 particles recovers the classical electric dipole-like absorption characterized by its $\nu^2$ dependence. Thus, the absorption in the classical regime ($\nu > E_c$) is insensitive to the particle size because the electrons no longer see the surface boundary in this frequency. We propose that the physical origin of the longstanding puzzle "the anomalous far-IR absorption by small metal particles" lies in the absorption due to the transitions of the electrons between the discrete energy levels in the quantum regime ($\nu < E_c$) on which the size-independent electric dipole-like absorption develops for $\nu > E_c$.

Figure Captions:

Figure 1    Far-infrared absorption coefficient of $f = 0.002$ Au55 particles dispersed in CsI at 10 K (black curve). Since the absorption coefficient is linear in $f$ for dilute system, the $\alpha_{Au55/CsI} - \alpha_{CsI}$ data has been rescaled to match that of the $f = 0.01$ $\alpha_{Au55/Teflon} - \alpha_{Teflon}$ (gray curve) from Ref [14]. The dashed curve is a fit to $0.11(v-10)^{0.8}$ and the dotted curve is the classical electric dipole absorption for $a = 5.3$ Å Au particle. Also shown is the absorption coefficient of pure CsI pellet at 10 K (light gray). The dip indicated as CsI is due to overcompensation in the process of subtraction of the modes present in CsI (see the text). The mismatches of the data in the overlapping frequency region are due to the change of the beam splitters.

Figure 2    The rescaled mid-infrared range absorption curve of $f = 0.002$ Au55 in CsI at 10 K. The numerous derivative-like modes occurred during the subtraction of the absorption coefficient of ligand/CsI sample from that of Au55/CsI (see the text and Ref. 15). However, the overall absorption baseline follows the $v^2$ behavior.

Figure 3    Comparison of the absorption coefficient of Au55 (solid gray line) to the classical electric dipole absorption coefficient for $a = 5.3$ Å Au particles at $f = 0.01$ (dotted curve) and for $a = 400$ Å Au particles at $f = 0.01$ with a constant 4 cm$^{-1}$ offset (dashed curve). The mismatches of the experiemntal data in the overlapping frequency region are due to the change of the beam splitters/detectors.

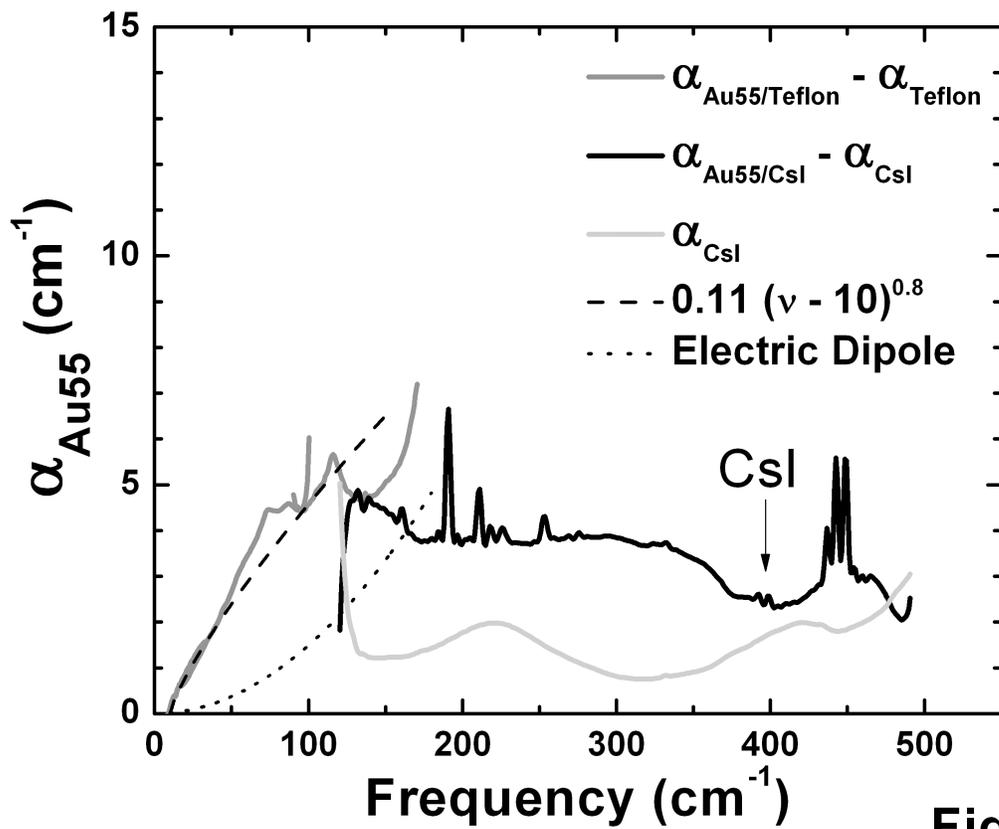

Figure 1

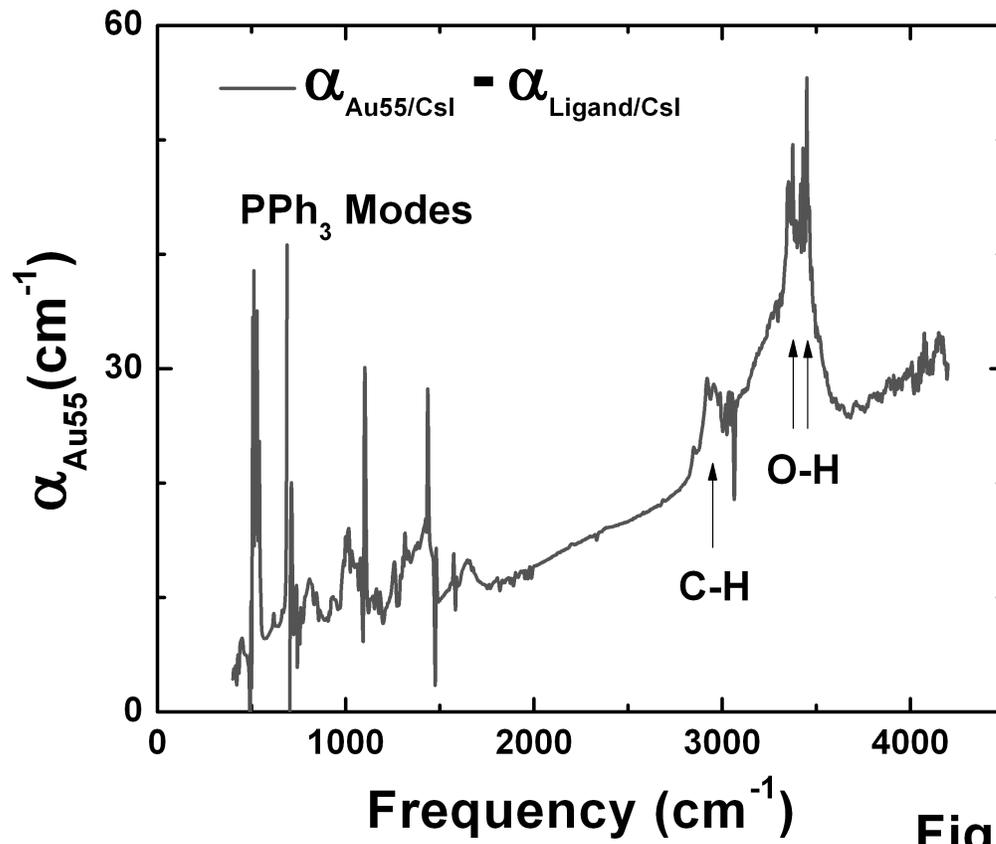

Figure 2

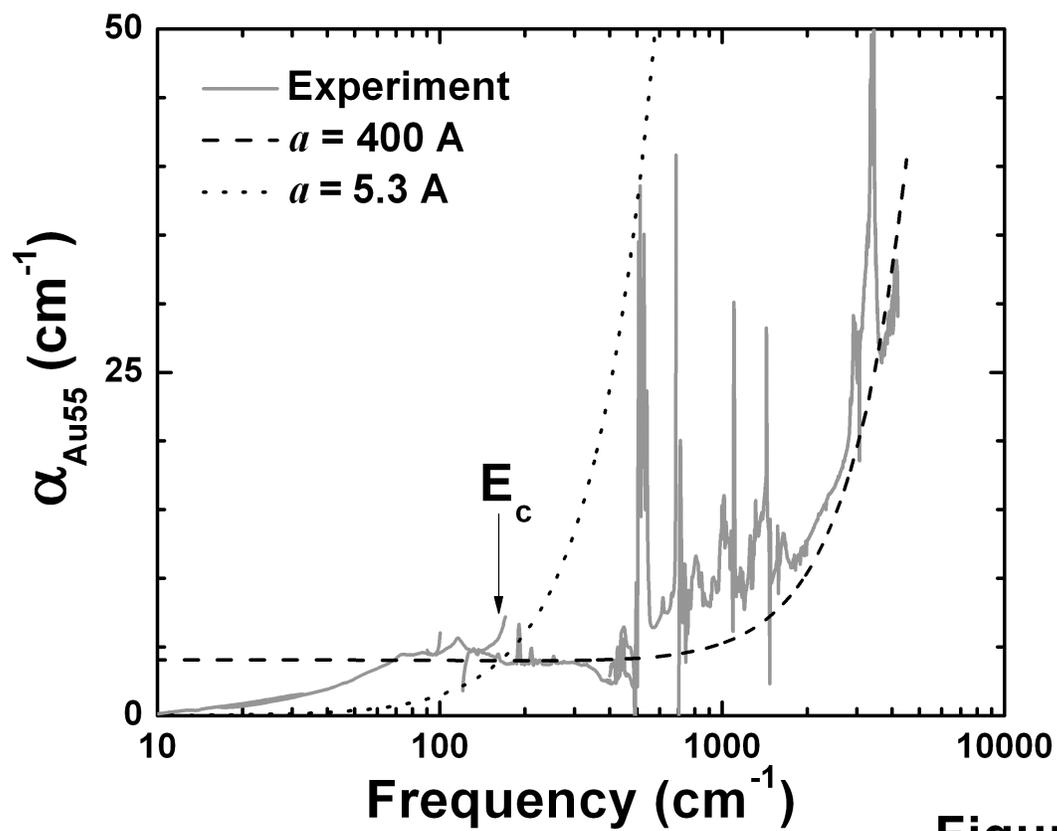

Figure 3